\documentclass{aa}
\usepackage{graphicx}
\usepackage{verbatim}
\usepackage{amssymb}
\usepackage{natbib}
\bibpunct{(}{)}{;}{a}{}{,}
\newcommand{\pas}{.\hskip-2pt$^{\prime\prime}$}
\titlerunning{Water maser's kinematics in massive YSOs}
\begin{document}
   \title{Kinematics of H$_2$O masers in high-mass star forming regions }

  \author{C. Goddi \inst{1}\fnmsep\inst{2} 
\and L. Moscadelli \inst{1} \and  W. Alef \inst{3} \and A. Tarchi \inst{1}\fnmsep\inst{4}
\and J. Brand \inst{4}\and M. Pani \inst{2}}

   \offprints{C. Goddi,\\\email{cgoddi@ca.astro.it}}

   \institute{INAF, Osservatorio Astronomico di Cagliari, Loc. Poggio dei Pini,
 Str. 54, 09012 Capoterra (CA), Italy
   \and
      Dipartimento di Fisica, Universit{\`a} degli Studi di Cagliari,
       S.P. Monserrato-Sestu Km 0.7, I-09042 Cagliari, Italy
         \and
             Max-Planck-Institut f{\"u}r Radioastronomie, Auf dem 
H{\"u}gel 69,
             D-53121 Bonn, Germany
          \and 
            Istituto di Radioastronomia CNR, Via Gobetti 101, 40129 Bologna, Italy 
             }

   \date{Received ``date'' / Accepted ``date''}

   \abstract{

We have conducted multi-epoch EVN observations of the 22.2~GHz water masers towards four high-mass star forming regions  (Sh~2-255 IR, IRAS~23139+5939, WB89-234, and OMC2). The (three) observing epochs span a time range of 6 months. In each region, the H$_2$O maser emission likely originates close (within a few hundreds of AU) to a forming high-mass YSO.
Several maser features ($\sim$~10) have been detected for each source and, for those features persistent over the three epochs, proper motions have been derived. The amplitudes of the proper motions are found to be larger than the range of variation of the line-of-sight velocities and in each of the observed sources the proper motion orientation seems to indicate an expansion motion.
 The gas kinematics traced by the 22.2~GHz H$_2$O masers is compatible with the shock-excited nature of water maser emission.

Three different kinematic  models (a spherical expanding shell, a Keplerian rotating disk, and a conical outflow) were fitted to the 3-dimensional velocity field of the detected maser features. The results of these fits, together with the comparison of the VLBI maps with the highest-resolution  images of the  sources in several thermal tracers,  suggest that the water maser features are most likely  tracing the inner portion of the  molecular outflows detected at much larger-scales.
 
   }

  \maketitle

\section{Introduction}

\begin{table*}
\centering
\begin{tabular}{cccccccc}
\multicolumn{8}{c}{\footnotesize {\bf Table 1:} Description of the observed sources} \\
& & & & & & & \\
\hline\hline
\multicolumn{1}{c}{Source} & &  \multicolumn{2}{c}{Coordinates (1950)} & & 
\multicolumn{1}{c}{$V_{\rm cloud}$} &  
\multicolumn{1}{c}{Distance} & \multicolumn{1}{c}{L$_{\rm bol}$} \\
\multicolumn{1}{c}{name} & &\multicolumn{1}{c}{RA (h m s)} &
\multicolumn{1}{c}{DEC ($^\circ \ ' \ ''$)} &  &
   \multicolumn{1}{c}{(km s$^{-1}$)} & \multicolumn{1}{c}{(pc)}
   & \multicolumn{1}{c}{($10^3 L_{\odot}$)}  \\
\hline
& & & & & & & \\
% AFGL 5142 & & 05 27 30.0 & +33 45 39.9  &  &  --4.4& 1800 &  4.0  \\
Sh~2-255 IR &  &  06 09 58.60 & +18 00  13.0   &    & 7.2$^a$   & 2500$^b$ &  9.0$^c$  \\
IRAS~23139+5939 & &   23 13 58.91 &  +59 39  06.0  &     &  --44$^d$ & 4800$^d$ & 25$^d$   \\
WB89-234 &  &   23 00 24.1 & +56 41 42.2  &    &--53.4$^e$ & 5800$^e$ &  10$^e$\\
OMC2 & &  05 32 59.9   &  --05 11 26.8  &     &  12$^f$ & 480  & 1.5$^g$ \\
& & & & & & & \\
 \hline
\end{tabular}
%\tablenotetext{a}{\citet{Mir97};}
%\tablenotetext{b}{\citet{Eva77};}
%\tablenotetext{c}{\citet{Mez88};}
%\tablenotetext{d}{\citet{Eva77};}
\begin{flushleft}
{\footnotesize  Note.-- $^a$ \citealt{Mir97}; $^b$ \citealt{Eva77}; $^c$ \citealt{Mez88}; $^d$ \citealt{Sri02}; $^e$ \citealt{Bra98}; $^f$ \citealt{Aso00}; $^g$ \citealt{Ren96}.}
\end{flushleft}
\end{table*}
%DARE LE REFERENCES DI QUESTI PARAMETRI????
%Mettere LE POSIZ. ASSOLUTE DEI MASERS DI REF.??

Comparing the present state of knowledge of the formation process of high-mass
(M $\geq$~10~M$_\odot$) and low-mass stars, one can appreciate how little is
known about high-mass star formation. 
High-angular ($\sim$1~arcsec) resolution observations over an extended range of wavelengths (from centimeter to Near-Infrared (NIR)) have made it possible to recognize different evolutionary classes of low-mass pre-main-sequence stars (from Class 0 to Class II; e.g. \citealt{And93}; hereafter AWB). On the other hand, for the high-mass counterpart, only a tentative protostellar evolutionary scheme presently exists, and it is often hard to distinguish between protostars and young Zero Age Main Sequence (ZAMS) stars.
This lack of knowledge is a direct consequence of difficulties in observing
the earliest evolutionary phases of massive Young Stellar Objects (YSOs), because
high-mass stars are rare, form on much shorter timescales ($\sim$~10$^{5}$~yr)  
than those required for low-mass formation, and the sites of massive star 
formation are found at much larger distances (typically several kiloparsecs), requiring high-angular resolution observations. 
High-mass stars  also spend all of their pre-main sequence phase deeply embedded in their 
natal molecular cloud (in fact they arrive on the main sequence while still 
accreting matter), suffering high extinction which prevents their detection at optical and (quite often) NIR wavelengths.

A powerful diagnostic tool to investigate the first stages of the formation of massive stars is provided by Very Long Baseline Interferometry (VLBI) observations 
%, at radio wavelenghts (where the medium is trasparent), 
of  the maser transitions of several molecular species, such as OH, H$_{2}$O, CH$_{3}$OH, observed in the proximity (within a distance $\leq$~100~AU) of the high-mass (proto-)stars.
In particular, for fast-moving 22.2~GHz water masers, multi-epoch VLBI observations, reaching  angular resolutions of $\sim$1~mas (corresponding to $\sim$1~AU at a distance of 1~kpc), allow accurate proper motions to be determined with time baselines as short as a few months. The measured tangential velocities, combined with the radial velocities derived via the Doppler effect, permit to obtain the 3-dimensional velocity distribution of the masing gas. This technique is presently the only one that has the potential to derive the gas kinematics in the close proximity  (few AUs) of the (proto-)star.

Recent VLBI observations have shown that 22.2~GHz water masers 
 are preferentially associated with
collimated flows of gas (jets) found at the base of larger-scale molecular
outflows (IRAS~05413-0104, \citealt{Cla98}; IRAS~20126+4104, \citealt{Mos00}; W3~IRS~5, \citealt{Ima00};  W75N-VLA1, \citealt{Tor03}). In a few cases, linear clusters (size 
$\sim$10-100~AU) of maser features have been interpreted in terms of accretion
disks (e .g., NGC 2071 IRS~1 and IRS~3, \citealt{Set02}; AFGL~5142, \citealt{God04}, hereafter Paper I).

Accretion disks were proposed originally to explain  Very Large Array (VLA) observations of water masers (e.g., Cepheus A, \citealt{Tor96}; W75N-VLA2, \citealt{Tor97}).
Successive Very Long Baseline Array (VLBA) observations (with a gain in resolution of $\sim10^3$) by \citet{Tor01a,Tor03}, however,  have revealed that the water maser features closer to the YSOs  are  distributed along perfectly circular arcs (radius 62~AU for Cepheus~A~R5, 160~AU for W75N-VLA2), which are also expanding (with velocities of 
10~km~s$^{-1}$ for Cepheus~A, 28~km~s$^{-1}$ for W75N-VLA2). The authors 
interpret these structures as being due to a spherical ejection of material 
from a YSO located at the center of the maser circle, demonstrating that clusters of water maser features close (within hundreds of AU) to the YSO may participate in  expanding motions driven by a wide-angle wind, rather than be subject to rotating motions.

Measurements of the maser feature proper motions are essential to distinguish between the different kinematic scenarios, allowing to clarify whether the maser clusters are expanding from the YSO (supporting the collimated or wide-angle wind interpretation) or rather rotating and contracting (in agreement with the disk model). 
So far, only a relatively small number ($\approx$10) of intermediate and/or high-mass YSOs have been studied with multi-epoch VLBI observations of the 22.2~GHz water masers.

\citet{Tor03} proposed also that  different kinematic
structures (spherical shells vs. collimated flows) traced by the water masers
represent different evolutionary stages of the YSO. In this view, in the
earliest evolutionary phases YSOs would emit wide-angle winds, whilst during
later phases the material ejection would become more collimated.
Therefore, these recent observational results open up the interesting perspective of using the 22.2~GHz water masers to derive information on the
evolutionary stage of high-mass YSOs. 

In Paper I we presented the results of  an European VLBI Network (EVN) multi-epoch study of the 22.2~GHz water masers  towards  the high-mass star forming region (SFR) AFGL~5142. In the following, we report the results of water maser EVN observations of  4 additional  high-mass YSO-candidates (Sh~2-255 IR, IRAS~23139+5939, WB89-234, OMC2).

Section~2 of this paper describes our multi-epoch EVN observations and gives
technical details of the data analysis. Section~3 presents the observational results. 
In Section~4, we investigate plausible kinematic models for interpreting the measured positions and velocities of the maser features, explore the physical environment of the observed sources as implied by the kinematic considerations, and discuss the way of  deriving an evolutionary pattern for the studied YSOs.
Finally, conclusions are drawn in Section~5.

%__________________________________________________________________
%

\section{Observations and data reduction}

The four high-mass YSOs were observed in the \(6_{16}-5_{23}\) H$_2$O maser line (rest frequency 22235.080~MHz) using the EVN at three epochs (June, September, and November, 1997). A description of the most relevant  properties (i.e., source name, coordinates, velocity of the ambient medium, distance and bolometric luminosity) of each source  is given in Table 1 .

An extensive description of the observations and of the data reduction process has already been provided in Paper I. Hence, here we  describe only the  observational parameters, summarized in Table~2, peculiar to the individual sources. 
For each  epoch, we used as phase-reference the peak intensity channel, which exhibits in all cases a simple spatial structure consisting of a single,  unresolved spot.   Due to the high water maser variability, the peak maser emission may vary in velocity (and position) from epoch to epoch. In particular, for the source Sh~2-255 IR the phase reference component of the third  observing epoch differs from the one used at the previous two epochs (see Table~2).
However, the maser components selected as phase reference were present at all the observing epochs, allowing us to confidently align maps of different epochs.
%For each source, both circular polarizations were recorded with a 2~MHz bandwidth, excepted WB234 and AFGL 5142, whose signal was contained in a 1~MHz bandwidth.

For all the four sources and the three observing epochs, we produced (naturally weighted) maps extended over a sky area of $(\Delta \alpha \  cos\delta \times \Delta \delta) \ 2''\times 2''$.
The CLEAN beam was an elliptical gaussian, with FWHM size and position angle varying with the epoch (see Col.~5 of Table~2). 
 The RMS noise level on the channel maps, $\sigma$, (reported in Col. 6) varies significantly, being close to the theoretical thermal value  for channels where no signal is detected and increasing up to one (or even two) orders of magnitude for channels with the strongest components.

Every channel map has been searched for emission above a conservative 
detection threshold, taken equal to the absolute value of the minimum in the map  corresponding to a multiple of the rms noise by a factor in the range 5--10. 
The detected maser spots have been fitted with two-dimensional elliptical Gaussians, determining  position, flux density, and FWHM size of the emission.  

Following Paper I, we define a ``maser feature'' as a group of maser spots detected in at least three contiguous channels, with a position shift of the intensity peak from channel to channel smaller than the FWHM size. 
%We verified that in all the cases most of the single-dish (or: total-power) flux was recovered in the imaged VLBI field of view. [VERIFICARE]

\begin{table*}
\centering
\begin{tabular}{cccccc}
\multicolumn{6}{c}{\footnotesize {\bf Table 2:} Summary of water maser observations} \\
& & & & &\\
\hline\hline
 \multicolumn{1}{c}{Source} &  \multicolumn{1}{c}{Epoch} &  \multicolumn{1}{c}{P. R. $V_{\rm LSR}$ } &\multicolumn{1}{c}{P. R. intensity}&\multicolumn{1}{c}{Beam size} & \multicolumn{1}{c}{ RMS noise}   \\
\multicolumn{1}{c}{} &  \multicolumn{1}{c}{}  & \multicolumn{1}{c}{(km s$^{-1}$)} &  \multicolumn{1}{c}{(Jy beam$^{-1}$)}&   \multicolumn{1}{c}{(mas)}   & \multicolumn{1}{c}{(mJy beam$^{-1}$)} \\
\hline
%& Oct 96 & --4.8 & 5.5 & 1.5$\times$1.0 pa= 70.2$^\circ$ & 20-60  \\
% AFGL 5142 & June 97 &--7.0 & 30.0 & 1.7$\times$1.4 pa= 76.7$^\circ$& 20-200  \\
%& Sep 97 & --4.8& 6.5 & 2.1$\times$0.9 pa= 54.9$^\circ$& 90-300 \\
%& Nov 97 & --4.8 & 7.6  &2.8$\times$1.2 pa= 59.7$^\circ$& 40-70 \\
& & & & \\
Sh~2-255 IR &  Jun 97& 11.6 & 41.8 & 1.1$\times$1.0 pa= 83.5$^\circ$ & 100-300 \\
 & Sep 97 &11.8 & 78.5  & 2.8$\times$1.2 pa= 50.6$^\circ$& 50-600 \\
& Nov 97 & 7.6 & 20.5 & 1.4$\times$1.0 pa= 86.5$^\circ$ &  20-100\\
& & & & \\
&  Jun 97 & --53.2 & 38.6 &1.6$\times$1.4 pa= --81.6$^\circ$ & 40-500 \\
IRAS~23139+5939 & Sep 97 &--53.3 & 247.0 & 2.3$\times$1.6 pa= --41.4$^\circ$& 40-4000\\
& Nov 97 & --53.1 & 240.0 & 1.5$\times$1.3 pa= 87.1$^\circ$ & 20-2000\\
&  & & & \\
& Jun 97& --51.1 & 15.0 &  1.5$\times$0.7 pa= 42.1$^\circ$ & 100-300 \\
WB89-234 & Sep 97 &--51.1 & 17.8  & 1.4$\times$0.9 pa= --61.9$^\circ$ & 200-300 \\
& Nov 97 & --51.1 & 32.4  & 1.1$\times$0.9 pa= --57.8$^\circ$ & 100-300\\
& & & & \\
&  Jun 97 & 18.0 & 10.2 & 2.4$\times$1.6 pa= --38.5$^\circ$ & 30-200 \\
OMC2  & Sep 97 &18.0 & 70.0 & 2.9$\times$1.4 pa= 43.9$^\circ$ & 50-600 \\
& Nov 97 & 18.0& 31.3 & 2.3$\times$1.1 pa= 54.2$^\circ$ & 100-900\\
& & & & \\
 \hline
\end{tabular}
\begin{flushleft}
{\footnotesize  Note.-- Cols.~3 and~4 give  the LSR velocity and the peak intensity of the  maser component selected as phase reference.}
\end{flushleft}
\end{table*}

%__________________________________________________________________ 
%
\section{Observational results}

\begin{table*}
\centering
\begin{tabular}{ccccccccccc}
\multicolumn{11}{c}{\footnotesize {\bf Table 3:} Maser feature parameters } \\
& & & & & & &  & & & \\
\hline\hline
\multicolumn{1}{c}{Source} & \multicolumn{1}{c}{Feature} & \multicolumn{1}{c}{$V_{\rm LSR}$} &
\multicolumn{1}{c}{$F_{\rm int}$} &  &
\multicolumn{1}{c}{$\Delta \alpha$} & \multicolumn{1}{c}{$\Delta \delta$} &
& \multicolumn{1}{c}{$V_{\rm x}$} & \multicolumn{1}{c}{$V_{\rm y}$} &
\multicolumn{1}{c}{$V_{\rm mod}$} \\
\multicolumn{1}{c}{name} & & \multicolumn{1}{c}{(km s$^{-1}$)} &
\multicolumn{1}{c}{(Jy)} &   &
   \multicolumn{1}{c}{(mas)} & \multicolumn{1}{c}{(mas)} &
   & \multicolumn{1}{c}{(km s$^{-1}$)} & \multicolumn{1}{c}{(km s$^{-1}$)} &
   \multicolumn{1}{c}{(km s$^{-1}$)} \\
\hline
& & & & & & &  & & & \\
Sh~2-255 IR  & 1        &  11.6 & 51.9 &  &   --147.8 (0.1) &  --100.69 (0.09) & &  8 (7)   &  --30 (5) & 31 (5) \\
 & 2        &  7.3 & 24.7 &      &   1.4 (0.1) &  --0.71 (0.05) & & 5 (6) &  --11 (3) &  12 (4) \\
 & 3        & 9.9 & 2.6 &  &   2.77 (0.08) &  0.39 (0.05) & &  &    &   \\
 & 4        &  5.6 & 5.0 &  &   63.1 (0.1) &  9.09 (0.05) & & {\it 7 (11)}$^{\dag}$ & {\it--14 (9)} &  {\it 16 (9) }    \\
 & 5        &  12.9 & 2.4 & &   --140.0 (0.2) &  --90.1 (0.2) & & {\it14 (15)} &  {\it--41 (12)} &   {\it43 (12)}   \\
 & 6        &  10.5  & 2.5  & &   --16.7 (0.2) &  --78.4 (0.2) & &  &   &  \\
 & 7        &  10.7 & 2.4 & &   1.5 (0.2) &  0.8 (0.2) & &  &    & \\
 & 8        & 12.5 & 11.5 & &   --149.2 (0.2) &  --102.2 (0.3) && 5 (7) &  --35 (8)& --35 (8)      \\
 & 9        & 9.1 & 13.9 & &    0.0 (0.2) &  0.0 (0.07) & & 0  &  0  & 0  \\
  & 10        &  6.7 & 4.9 & &  8.5 (0.2)    &  1.6 (0.2)    & &  & & \\
 & 11        &  5.8 & 0.6 &  &  --132.8 (0.2) &  --181.0 (0.2) & &  &   &  \\
  & 12       &  6.4 &4.0 &  &   --135.6 (0.1) &  --180.75 (0.05) & & &  &   \\
 & 13       &  7.1 & 2.5 &  &  81.3 (0.1) &  18.40 (0.05) & & &  &   \\
& 14& 8.5& 2.7 & & --158.3 (0.1) & --187.61 (0.05)&  & & & \\
& & & & & & &  & & & \\
IRAS 23139+5939 & 1& --53.2 & 199.6 && --2.81 (0.08) &  2.45 (0.06)&& {\it--13 (15)}& {\it 11 (9) }&{\it 17 (12)}  \\
& 2 & --54.7& 7.1 & &0.00 (0.05)& 0.0 (0.05)& & 0 & 0& 0 \\
& 3& --50.5& 4.6 & & --16.82 (0.09)& --0.24 (0.07)& &{\it --40 (14)} &{\it 17 (18)} & {\it 44 (15)}\\
& 4 & --52.9 & 4.8 & & --67.3 (0.1) & --226.6 (0.1)&  &{\it 97 (17)} &{\it --57 (21)} &{\it 112 (18)}\\
& 5 &--46.6 & 4.2& &--153.78 (0.08) & --183.4 (0.1)& & & & \\
& 6 & --46.7 & 1.2 & &--11.70 (0.08) & --20.7 (0.1)&  & & & \\
& 7& --47.3 & 0.9& &14.2 (0.2) & --18.2 (0.1)&  & & & \\
& 8 & --57.7& 0.7& & --3.7 (0.1 )& 1.6 (0.1)&  & & & \\
& 9 &--46.6 & 0.8& & 24.49 (0.08)& 94.0 (0.2)&  & & & \\
& & & & & & &  & & & \\
WB89-234 & 1 & --51.0& 30.4& & 94.29 (0.06)& 18.21 (0.04)& & 11 (4) & --20 (5)& 23 (5) \\
& 2 &--54.9 & 6.6 & & 0.00 (0.08)&  0.00 (0.05)& &  0 &0 &0 \\
& 3&--55.9 & 4.7 && 5.35 (0.07)& --15.46 (0.04)& & 10 (5)& --31 (2)&  33 (3) \\
& 4& --50.1 & 4.6 & & 32.42 (0.07)& 9.11 (0.03)&& {\it 34 (16)} & {\it --37 (5) } &{\it 50 (11) } \\
& 5 & --57.0& 4.0& &3.6 (0.3) & --16.41 (0.09)&  & 0.2 (25)& --65 (8)& 65 (8)\\
& 6& --56.0& 2.1& & --12.04 (0.07)& --13.32 (0.03)&  & & & \\
& 7& --58.4& 1.4& & 5.4 (0.2)& --17.15 (0.04)&  & & & \\
& 8 & --55.9& 3.8& &5.88 (0.06)& 20.31 (0.04)& &  & & \\
& 9& --52.4 &1.6 & &5.06 (0.07) & 18.24 (0.04)& & &   \\
& 10& --53.8& 1.5& &--1.69 (0.04) & 62.94 (0.02)&  & & & \\
& 11& --53.7& 1.4& & --1.19 (0.08)& --9.65 (0.05)&  & & & \\
& 12& --60.4 & 1.6& & 44.44 (0.03)& 13.73 (0.03)&  & & & \\
& & & & & & &  & & & \\
OMC2& 1& 17.6& 115.5& & 0.0 (0.1)& 0.00 (0.09)&  & 0& 0&0 \\
& 2& 11.9& 2.3& & 172.7 (0.1)& 89.29 (0.09)&  & & & \\
& 3& 8.5& 2.3& &158.5 (0.1) & 75.2 (0.1)&  & {\it 12 (3)}& {\it 13 (3)}& {\it 18 (3)} \\
& & & & & & &  & & & \\
 \hline
\end{tabular}
\begin{flushleft}
{ \footnotesize  Note.-- Col.~1 gives the source name;  for each identified feature Col.~2 reports the  label number; Cols.~3 and ~4  the line-of-sight velocity  and  the integrated flux density   of the highest-intensity channel; Cols.~5 and ~6  the (RA and DEC) relative positional offsets (with the associated uncertainties in parentheses); Cols.~7, ~8, and ~9  the projected components along the RA and DEC axes, and the absolute value of the derived proper motions.

${\dag}$ The italic character is used to indicate tentative values of proper motion components for  features observed either at only two epochs or with a very large uncertainty in the  direction of motion.}
\end{flushleft}
\end{table*}
In this section we present the VLBI maps of the observed water maser sources. 
From a comparison between these VLBI maps and those, at lower resolution, available in the literature, we can obtain further insights on the nature of the sources responsible for the water maser excitation.
VLA maps of the 22~GHz water maser emission were available for all the sources.

We have tentatively derived an absolute position for the EVN 22.2~GHz maser maps when the VLA and EVN strongest features were found to emit at the same LSR velocity, and under the assumption that the most intense maser emission emerges always from the same position. In doing so, the derived absolute position of the VLBI reference feature has an error  equal to (or greater than) the VLA positional uncertainty.

Table~3 reports the parameters of the identified water maser features. The procedure to derive (relative) positions, proper motions, and associated errors is described in detail in Paper I. In particular, we note that for each source the feature selected to refer positions and velocities to, is choosen using the criterion of increasing the amplitude of the relative proper motions, hence maximizing their SNR. The selected reference maser feature may have its own (absolute) proper motion, resulting in arbitrary offsets in the proper motions for the other features.

%____________________________________________________________________________
%
\subsection{Sh~2-255 IR}

The high-mass star forming complex Sh 2-255 IR, at a distance of 2.5 kpc from the Sun \citep{Eva77}, is located between the two HII regions S255 and S257.
VLA observations at 5~GHz (2\pas6 beam) detected three compact HII regions  associated with Sh~2-255 IR (Sh~2-255-2a, -255-2b and -255-2c), each one corresponding to a ZAMS star of spectral type B1 \citep{Sne86}. Successively, higher-resolution VLA observations at 15~GHz (beam $\approx$~0\pas5)  have revealed a continuum emission (flux density $\approx$~1.97~mJy) coincident with Sh~2-255-2c \citep{Ren96}.

\begin{figure*}
\centering
\includegraphics[width=14cm]{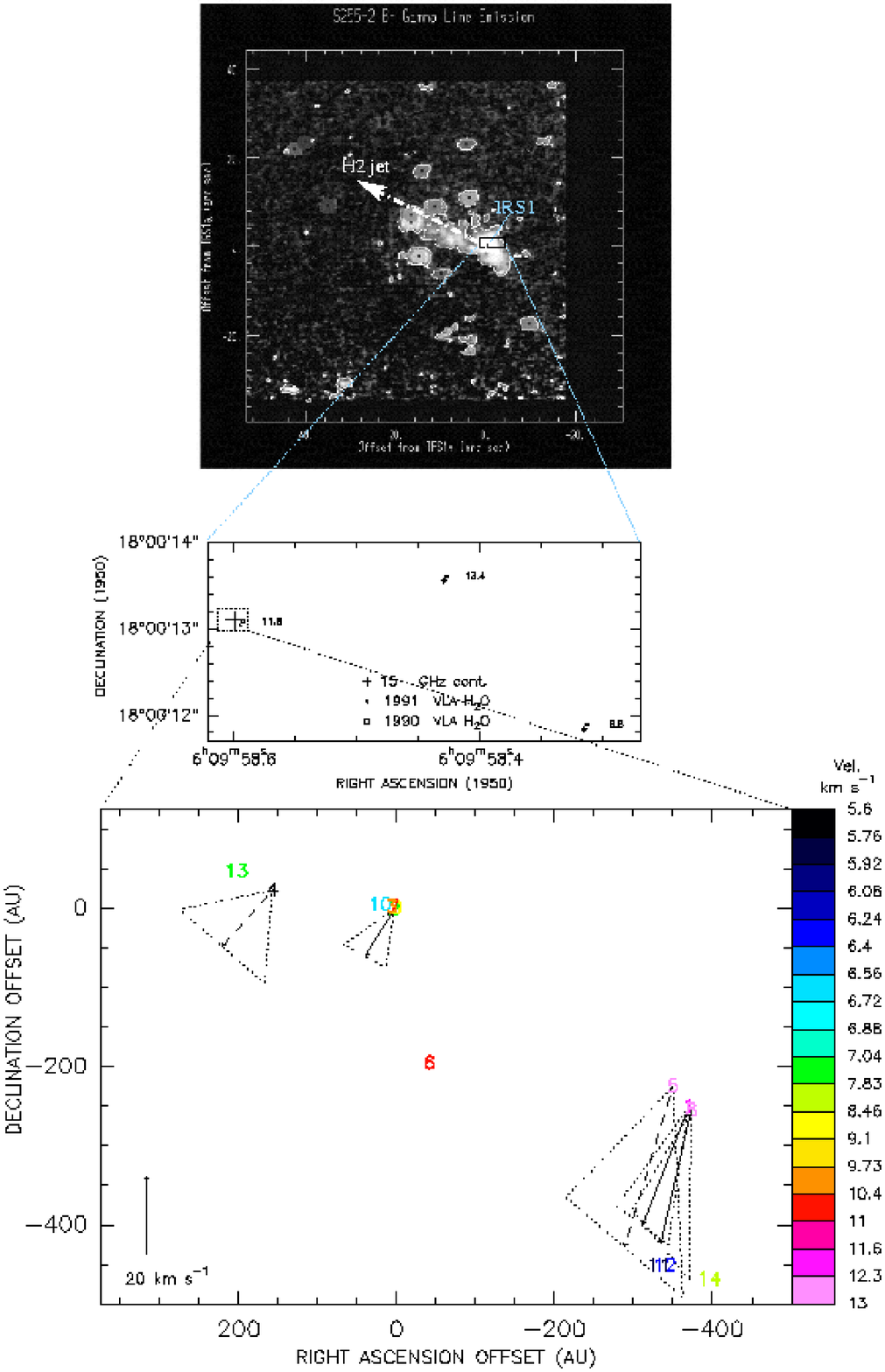}
\caption{Sh~2-255 IR. ({\itshape Upper panel}) Br$\gamma$ line minus continuum emission \citep{How97};  the white-dashed arrow indicates the proposed orientation of the jet observed also in the H$_2$ 2.12~$\mu$m emission (P.A.$\approx$ 67$^\circ$); the origin of the map coincides with IRS1 at  $\alpha$(1950)= $06^h 09^m 58^s.6$, $\delta$(1950)= $18^\circ  00' 13''$. 
({\itshape Middle panel}) Positions and the LSR velocities of the 22.2 GHz H$_2$O maser features detected with the VLA in 1990 ({\itshape open squares}) and 1991 ({\itshape filled dots}) by R. Cesaroni (priv. com.); the cross  indicates the positional uncertainty of the VLA 15~GHz continuum source \citep{Ren96}. 
({\itshape Bottom panel}) VLBI 22~GHz water maser features,  identified with the label number given in Col.~2 of Table~3. Different colours are used to distinguish the line-of-sight velocities of the features, according to the colour-velocity conversion code shown on the right-hand side of the panel. 
The arrows indicate the measured proper motions (whose amplitude scale is given at the bottom of the panel), with dashed lines used in case of features detected at only two epochs; the dotted triangles drawn around the arrows represent the  orientation uncertainty of the proper motions; dotted lines without error triangles are finally used in the case that  the signal-to-noise ratio of the proper motion is so small that the direction of motion is very uncertain. 
Positions (in AU) and proper motions are relative to the feature labeled ``9'', with absolute position: $\alpha$(1950)= $06^h 09^m 58^s.60$, $\delta$(1950)= $18^\circ 00'$ 13\pas2.}
\label{s255_maps}
\end{figure*}

Sh~2-255 IR was imaged by \citet{Mir97} at NIR and mm wavelengths, revealing a  cluster of 50 NIR sources, associated with Herbig Haro-like objects, IR H$_2$ jets and molecular outflows.
The reddest source in the region, NIR 3,  coincident with the radio source Sh~2-255-2c, is interpreted in terms of a YSO powering an infrared H$_2$ jet (aligned at P.A.$\approx$67$^\circ$) and a  CO  compact outflow (size $\leq$~1$'$),   approximately oriented parallel to the infrared-jet axis (\citealt{Mir97}; see  their Fig. 9).
High-resolution (beam $\approx$ 1\pas5) NIR observations of the H$_{2}$ 2.122$\,\mu$m and of the Br$\gamma$ hydrogen recombination lines by \citet{How97} confirm the presence of  an ionized jet (shown in the upper panel of Fig.~\ref{s255_maps}), likely originating from IRS1 (NIR 3  in the notation of \citealt{Mir97}), a  YSO candidate detected both at near- and mid-infrared wavelengths.
%Bolometric luminosity 2.6 10$^4$ L0 \citep{Jaf84} [beam 50 as]: comprende la S255N, quindi non {\`e} attendibile, meglio quello ad alta risoluzione:
%Bolometric luminosity= 0.9 x 10$^4$ L0 (Mezger 88) [beam 11-30 as]
%mm continuum: S1300= 2.2 Jy, MRT(Pico Veleta)= 11'' beam (Mezger 88)

%Molecular cloud velocity: 7.17 km/s, from the CO emission citet{Mir97}
%Methanol masers (Minier00) are coincident in position with our absolute positions of water masers [ma nn so l'accuratezza delle posizioni del metanolo, perch{\`e} Minier ha fatto osservaz. VLBI, ma le posizioni asolute sono da private communication: le {\`e}posizioni VLBI sono relative] 
%S255 IR  is also associated with CH$_3$OH \citet{Min00}, OH \citep{Tur71} and H$_2$O \citep{Lo75} masers.
%The radio to infrared emission properties of Sh~2-255-2c/NIR 3 suggest the presence of a high-mass YSO of ZAMS spectral type B1. 
22.2~GHz water maser VLA observations (unpublished data 
kindly provided by R. Cesaroni) at two epochs (March 1990; August 1991) reveal 
three emission centers over a region of a few arcseconds (see middle panel of Fig.~\ref{s255_maps}), with the strongest maser feature ($\approx$30~Jy) found to be coincident in position with Sh~2-255-2c. 

The lower panel of Fig.~\ref{s255_maps} shows the positions 
and velocities of the 22.2~GHz maser features as derived by  our multi-epoch EVN observations.
Maser features show an elongated spatial distribution of 
size 
$\approx$700~AU whose major axis is oriented approximately parallel to the ionized jet axis observed at arcsec-scale using the Br$\gamma$ and H$_{2}$ 2.12$\,\mu\,m$ lines. The line-of-sight velocity dispersion in the molecular outflow agrees well with the  LSR velocity  dispersion of VLBI maser features.
All the measured {\em relative} proper motions are approximately perpendicular to the jet axis, and have amplitudes (in the range \ 10 -- 40~km~s$^{-1}$) large compared to the spread of radial 
velocities ($\approx$7~km~s$^{-1}$).

%____________________________________________________________________________

\subsection{IRAS~23139+5939}

%[This source is included in the list of high-mass protostellar objects that Beuther et al.(2002) and Sridharan et al.(2002) studied with angular resolution of $\sim$1~arcsec over an extended range of wavelenghts (from cm to sub-mm) using both molecular line and continuum tracers. 

A multi-wavelength study of IRAS~2313+5939 (at a distance of $\approx$ 4.8~kpc) conducted by \citet{Sri02}  and \citet{Beu02b} have revealed a mid-infrared source (positional accuracy $\approx$ 5$''$), coincident in position  with a Plateau de Bure
(PdB) 1.2~mm continuum source (flux  density $\approx$ 2.3~Jy, for an observing beam $\approx$ 5$''$) and with a faint 3.6~cm continuum source detected at two different epochs with the VLA (1992,  0.6~mJy flux  density  with a 0\pas3~beam, \citealt{Tof95}; 1998, 1.4~mJy flux density with a 0\pas7~beam, \citealt{Sri02}). The cm-source has been interpreted in terms of (optically thin) free-free emission but no spectral index information was provided to support this interpretation. Nevertheless, these observations strongly suggest the presence of a massive YSO in this region. Its bolometric luminosity, $\approx 2.5 \times 10^4$ L$_{\odot}$  \citep{Sri02}, would correspond to a ZAMS star of spectral type B0 \citep{Pan73}. 
Unfortunately, the low angular resolutions of the previously listed observations do not allow to  establish  whether the cm to infrared  wavelength emission is powered by a single YSO or is the cumulative result of several distinct sources.

\begin{figure*}
\centering
\includegraphics[width=\hsize,trim= 0cm 0cm 0cm 3cm,clip]{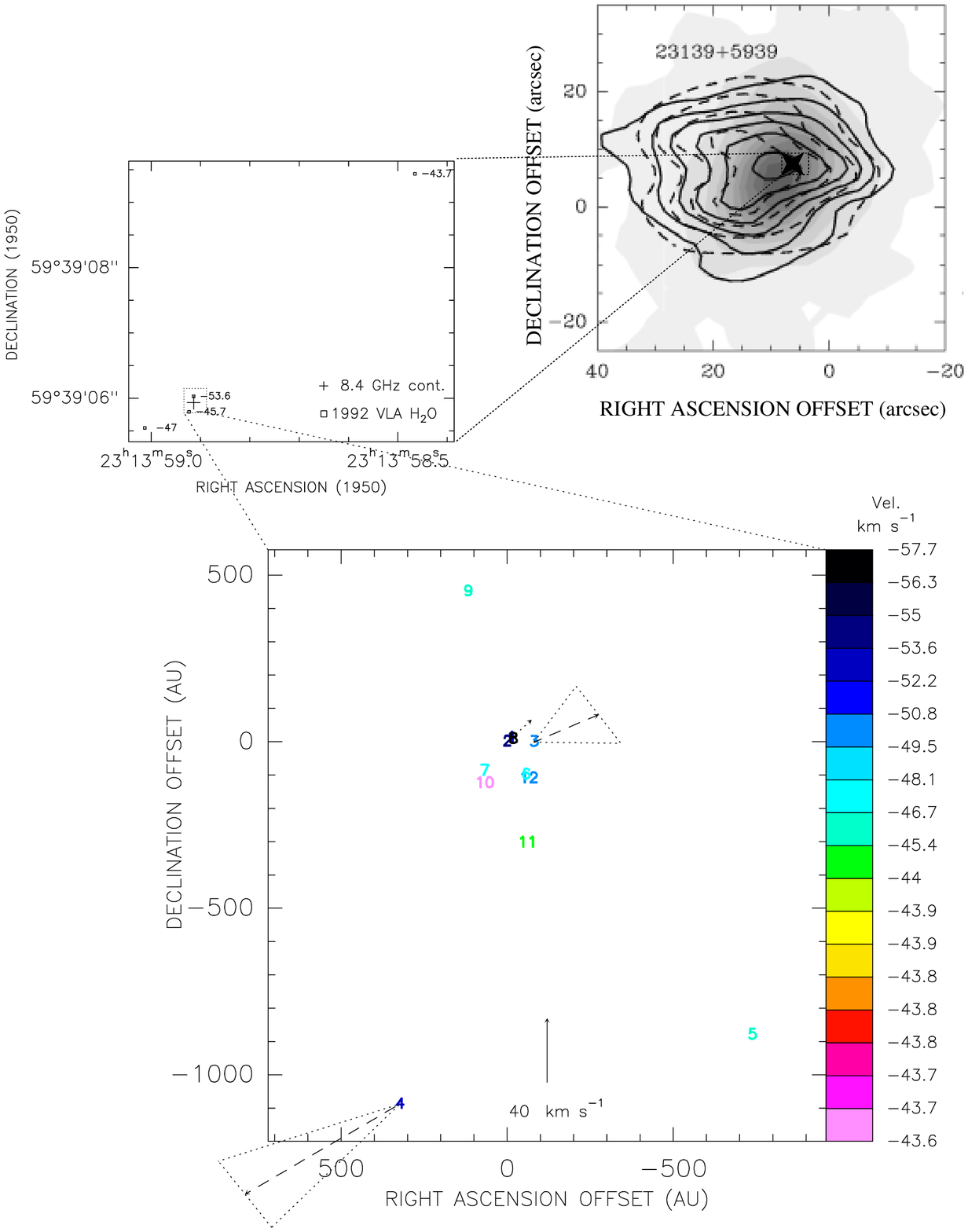}
\caption{IRAS~23139+5939. ({\itshape Upper panel}) Blue lobe ({\itshape solid lines}) and red lobe  ({\itshape dashed lines}) of CO 2-1 emission mapped with the PdB by \citet{Beu02b}, overlapped with  the  PdB 1.2~mm continuum emission  ({\itshape grey scale}); 
the axes show offsets in arcsec from the absolute  IRAS-position:  $\alpha$(1950)=$23^h 13^m 58^s.016$, $\delta$(1950)= $59^\circ 38'$ 59\pas890 \citep{Sri02}; the filled star indicates the position of the infrared source.
({\itshape Middle panel}) Positions ({\itshape open squares}) and the LSR velocities of 22.2~GHz H$_2$O maser features detected with the VLA in 1992 by \citet{Tof95}; the cross  indicates the positional uncertainty of the 8.4~GHz continuum source. ({\itshape Bottom panel}) VLBI map of the 22.2~GHz water maser features (see the caption of Fig. \ref{s255_maps});  the axes show offsets in AU from the absolute position: 
 $\alpha$(1950)= $23^h 13^m 58^s.914$, $\delta$(1950)= $59^\circ  39'$ 06\pas03, corresponding to the feature with label number ``2''.}
\label{i2313_maps}
\end{figure*}

A CO  outflow is detected towards IRAS 23139+5939 \citep{Wou89a, Beu02b}, nearly oriented along the line-of-sight and not spatially resolved at single-dish angular resolutions (beam $\approx 11''$) (Fig.~\ref{i2313_maps}, upper panel).

The 22.2~GHz water masers were observed using the VLA by \citet{Tof95} (0\pas1~beam) and \citet{Beu02a} (0\pas4~beam).
\citet{Tof95} detected 4 maser features distributed at the center of the molecular outflow (with line-of-sight velocities in agreement with the blue-shifted lobe): three of these are clustered around the 8.4~GHz continuum source, while the fourth is offset from it by 5$''$  (Fig.~\ref{i2313_maps}, middle panel). \citet{Beu02a} detected only the maser features associated to the radio continuum source but a precise correspondence between the features at the two VLA epochs is impossible because of the different accuracy in absolute positions (0\pas1, \citealt{Tof95}; 1$''$, \citealt{Beu02a}).

Fig.~\ref{i2313_maps} (lower panel) shows positions and velocities of the 22.2~GHz
maser features as derived from our multi-epoch  EVN observations .
Most of the 22.2~GHz emission emerges from a  cluster of
maser features whose diameter is $\approx$200~AU,  adopting a kinematic distance of 5~kpc. 
 Most of the VLBI features have line-of-sight velocities blue-shifted with respect to the quiescient gas velocity ($\approx$~--44~km~s$^{-1}$) and distributed over the whole velocity range of the blue-shifted lobe of the CO molecular outflow, $\Delta v = (-59,-47)$~km~s$^{-1}$ \citep{Beu02b}.
Although only a  small number of proper motions is derived, they 
are suggestive of a general expansion motion with (relative) velocities of tens of km~s$^{-1}$ (see Sect. 4).

%____________________________________________________________________________

\subsection{WB89-234}
\begin{figure*}
\centering
\includegraphics[width=\hsize, trim = 0cm 2cm 0cm 9cm,clip]{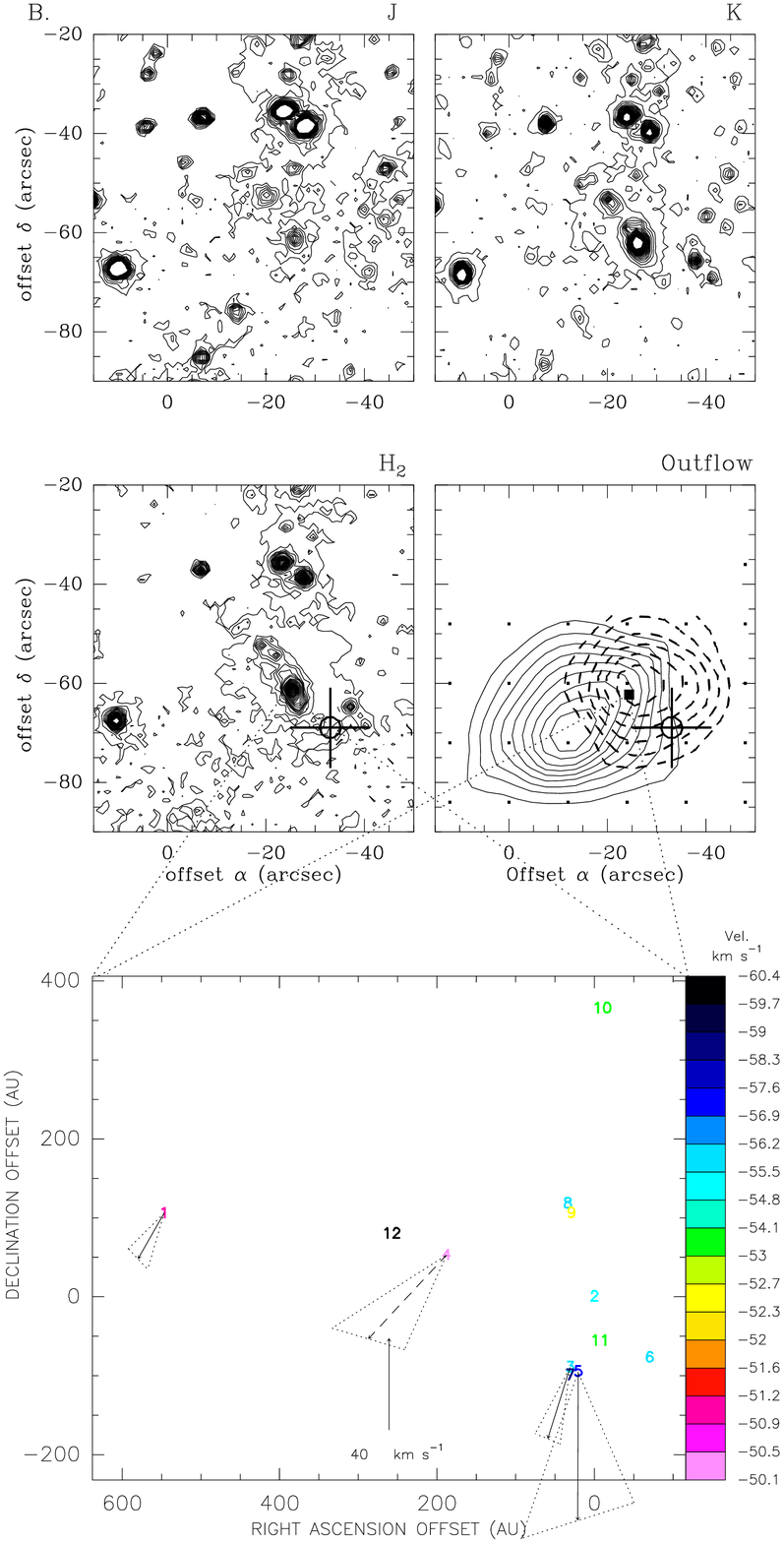}
\caption{WB89-234. ({\itshape Upper panels}) TIRGO H$_2$~2.12~$\mu$m image ({\itshape left}); blue ({\itshape dashed lines}) and red  ({\itshape solid lines}) lobes  of $^{12}$CO(1--0) wing emission \citep*{Bra98} ({\itshape right}); the big  crosses and the filled squares indicate the positional uncertainties of the H$_2$O maser observed respectively with Effelsberg \citep*{Bra98} and the VLA (Brand \& Wouterloot, unpublished); the axes show offsets in arcsec from the absolute  IRAS-position:  $\alpha$(1950)=$23^h 00^m 27^s.0$, $\delta$(1950)= $56^\circ 42' 45''$.
({\itshape Bottom panel}) VLBI map of the 22.2~GHz water maser features (see the caption of Fig. \ref{s255_maps}); the origin of the map corresponds to the position of the feature ``2'', at: $\alpha$(1950)=$23^h 00^m 24^s.150$, $\delta$(1950)= $56^\circ 41'$ 42\pas3.}
\label{wb234_maps}
\end{figure*}

WB89-234 (at a distance of 5.8~kpc) was first detected by
\citet*{Wou89} during a survey of CO emission towards IRAS sources in
SFRs and subsequently studied in detail by \citet*{Bra98}. Fig. ~\ref{wb234_maps} (upper panels) shows that the H$_{2}$O maser
(whose absolute position is derived from VLA observations; Brand \& Wouterloot, unpublished)
is found at the center of a CO bipolar outflow and coincides with a 
NIR source (imaged in J, H, K bands, and in the H$_{2}$~2.12~$\mu$m line), 
which
probably identifies the embedded exciting YSO. The associated water maser has been regularly monitored with the  Medicina 32--m radiotelescope during the last ten years (from 1994 to 2004), revealing a velocity range of emission roughly constant between --60 and --50~km~s$^{-1}$ (within the velocity range of the CO outflow), with the strongest component usually at $\approx$~--51~km~s$^{-1}$. 

The 22.2~GHz maser features 
detected in our EVN maps (Fig. ~\ref{wb234_maps}, bottom panel)
present a line-of-sight velocity dispersion consistent with the Medicina observations (with the intensity-peak at --51.1~km~s$^{-1}$) and are distributed across an 
area of diameter $\approx$900~AU. The maser spatial distribution shows no preferential direction of elongation. All the measured {\em relative}
proper motions have similar northwest-southeast orientation (close to the 
direction of the molecular outflow), with amplitudes (in the range \ 20 -- 
60~km~s$^{-1}$) large compared with the spread of line-of-sight velocities 
($\approx$10~km~s$^{-1}$). 

%____________________________________________________________________________

\subsection{OMC2}

\begin{figure*}
\centering
\includegraphics*[width=\hsize,trim= 0cm 0cm 0cm 5cm,clip]{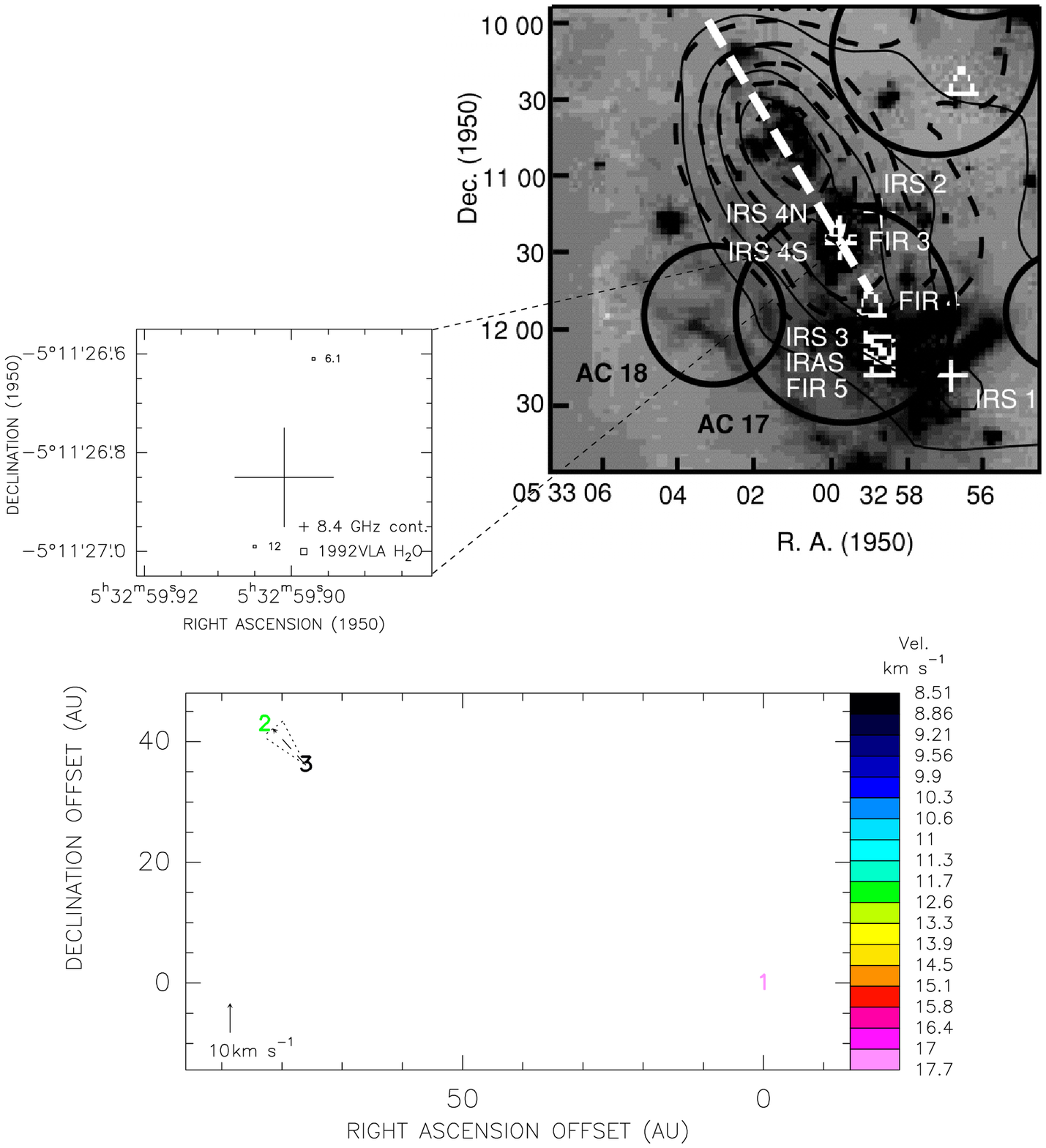}
\caption{OMC2. ({\itshape Upper panel}) Blue wings ({\itshape solid lines}) and red wings  ({\itshape dashed lines}) in the HCO$^+$ line tracing the outflow \citep{Aso00}, superimposed on the image of H$_2$ emission ({\itshape grey scale}) \citep{Yu97}; the white dashed-line indicates the elongation axis of the outflow; the positions of IRS4 and FIR 3 are also shown.
({\itshape Middle panel}) Positions ({\itshape open squares}) and the LSR velocities of 22~GHz H$_2$O maser features detected with the VLA in 1992 by \citet{Tof95}; the cross  indicates the positional uncertainty of the 8.4~GHz continuum source. 
({\itshape Bottom panel}) VLBI  map of the 22.2~GHz water maser features (see the caption of Fig. \ref{s255_maps}), with positions relative to the reference feature ``1''.}
\label{omc2_maps}
\end{figure*}

OMC2, located at a distance of about 480~pc, is one of the most active sites of star formation in the solar neighborhood. A cluster of young infrared sources \citep {Jon94} and several submillimeter continuum sources \citep {Mez90,Chi97} are found in this region, and 3.6~cm  continuum VLA observations ($\approx8''$~beam) have revealed an unusual concentration of class 0 protostars  \citep {Rei99}. One of these radio sources, labeled VLA~11 \citep {Rei99}, is coincident in position with a bright 1.3~mm source, FIR 3 \citep{Mez90}, and an infrared source, IRS 4 (detected at several different wavelengths from 1 to 100~$\mu$m;  \citealt{Pen86}), and is also associated with a weak 22~GHz water maser \citep{Gen79}. Besides, VLA~11 is found at the center of a group of bright near-infrared H$_2$ bow shocks (flow J of \citealt{Yu97}) and of a CO and HCO$^+$ outflows \citep{Fis85,Aso00}. The molecular outflows are nearly oriented along the line-of-sight (with the red and the blue lobes spatially overlapped),  but unlike the CO outflow, the HCO$^+$ outflow is elongated and well correlated with the axis of the H$_2$ jet (P.A. $\approx$~30$^\circ$; Fig.~\ref{omc2_maps}, upper panel). 

High-resolution observations ($\approx$0\pas1 and 0\pas5~beam, respectively), conducted with the VLA at 8.4~GHz \citep{Tof95} and 15~GHz \citep{Ren96},  revealed a faint ($\approx$2.5~mJy), slightly extended (0\pas8) continuum source, coincident in position with the sources VLA\,11/FIR\,3. The radio continuum source is found to be extended along a North-South direction (P.A. 17$^\circ$) and it is interpreted  as optically thin free-free emission from an ionized wind \citep{Ren96}, compatible with an intermediate-luminosity ZAMS star (spectral type between B2 and B3).

Using the VLA, \citet{Tof95} detected two  22~GHz maser features, placed  respectively northward and southward of the 8.4~GHz  continuum source, and separated by $\approx$0\pas5 (Fig.~\ref{omc2_maps}, middle panel). 

%The precise nature of the cm source is not well established. \citet{Tof95} exclude that its flux density (2.4 mJy) is due to emission from dust or from the embedded YSO itself and \citet {Rei99} suggest that it is probably due to free-free emission from shocks in radio jets. This interpretation is consistent with the fact that the cm source by \citet{Tof95} is nearly extended along a N-S direction, supporting the jet hypothesis.
%Lbol = 1500 L0 da reng96, diverso daTof95=200 L0
%mm: S1300= 2.6 Jy; MRT(IRAM)= 11'' beam (Mezger 90)

Only three maser features have been detected in our EVN maps towards this source and only one ``tentative'' proper motion has been derived for the feature with label number ``3'' (Fig.~\ref{omc2_maps}, bottom panel).
In this case a precise association between the VLA and the VLBI emission is impossible, since  no correspondence is found between the line-of-sight velocities of the strongest VLA and VLBI spots.
The three features are aligned along a direction similar to the orientation of the measured proper motion. This direction is also similar to the orientation of the large-scale molecular outflow seen in the  H$_2$ and CO emission. 

%__________________________________________________________________
%
\section{Discussion}
\subsection{Kinematics of the masing gas}

%DISCUSSIONE GENERALE SULL{\'I}MPOSTAZIONE DELL{\'A}NALISI
In Sect.~3, for each source, we have presented multi-epoch VLBI maps of the water masers and compared them with  previous observations in different thermal tracers.

Several concentrations of H$_{2}$O maser features are found in each of the observed sources within 1000~AU of the YSO position, which itself is indicated by unresolved radio and/or millimeter continuum sources.
%Two sources show almost elongated spatial distributions  of water maser features (Sh~2-255 IR and OMC2), whilst the remaining two (IRAS 23139+5939 and WB89-234) have more clustered distributions.
%Linear distributions of water masers could be simply explained with collimated outflows and/or edge-on keplerian disks.
Clustered distributions of maser features have been  interpreted, depending on the geometrical shape (curved or linear) and the measured proper motions, in terms of expanding spherical shells (representing wide angle winds) \citep{Tor01a,Tor03} and/or jet/disk systems (\citealt{Set02}; Paper I).

There is no clear evidence of spherical structures in the observed water maser distributions.
IRAS\,23139+5939 shows the most clustered distribution of maser features. Since the spherical geometry might be hidden by projection effects onto the plane of the sky, we constructed a simple kinematic model of spherical expansion, which assumes that  the detected maser features  belong to the surface of a sphere of given radius and move outward   with a velocity increasing with radius.  This model was applied to IRAS~23139+5939,  but the best fit solution was not physically plausible, since the maser features sampled a very small fraction of the sphere's surface, that moreover was centered well outside the maser region. Hence, we conclude that a spherical expansion model such as that proposed by \citet{Tor01a,Tor03} for the high-mass YSOs Cepheus A-R5 and W75N-VLA2 does not apply to the  distribution of maser features seen towards our sources. 
%[Anyway, one should note that the condition of spherical geometry  is not essential for the interpretation of maser clusters as excited by wide-angle winds. In principle, non-circular clusters of maser features  might result when a wide angle wind propagates across a not isotropical ambient medium, which could fragment the originally spherical wave front.]

The current theory of star formation foresees that two main kinds of gas motion should be found in the proximity of a forming star: 1) rotation and contraction, under the influence of the gravitational field of the central YSO (resulting in the creation of an accretion disk); 2) expansion, driven by either wide-angle or collimated winds (resulting in the formation of a jet/outflow system). 
We have tested the expansion vs rotation scenario by fitting the measured positions and velocities of the detected maser features with two simple, mutually exclusive kinematic models: \ 1) a Keplerian disk;  \ 2) a conical outflow. 
Since few maser features have been detected and only few accurate proper motions have been measured  towards the studied YSOs, we warn that the  results  obtained from these model fits should be considered only as a test for the physical plausibility of the kinematic structures traced by water masers in the proximity of the  massive YSOs,  without  any claim of conclusiveness.

The free parameters of the Keplerian disk model are: the sky-projected 
coordinates, $\alpha_{\rm d}$ (R.A.) and $\delta_{\rm d}$ (Dec), of the YSO (at the
disk center); the position angle, $P_{\rm d}$, and the inclination angle with the
line-of-sight, $i_{\rm d}$, of the disk axis; the YSO mass, $M_{\rm YSO}$.
The conical outflow model assumes that the gas moves along the directrices of a conical surface, with a velocity increasing with the distance from the cone 
vertex following the Hubble law: \ {\bf v} = $a_{\rm c}$ \  {\bf r}, where $a_{\rm c}$ is the velocity gradient. The parameters of the conical flow are: the 
sky-projected coordinates, $\alpha_{\rm c}$ (R.A.) and $\delta_{\rm c}$ (Dec), of the
YSO (at the cone vertex); the position angle, $P_{\rm c}$, and the inclination 
angle with the line-of-sight, $i_{\rm c}$, of the cone axis; the opening angle of the cone, $\theta_{\rm c}$; the velocity gradient $a_{\rm c}$.

The  parameters of the best fit models are reported in Table~4.
The best-fit solution is found minimizing the $\chi^{2}$ given by the sum of the square differences between the  3-dimensional, relative,  model and measured  velocities.
%The model fits were applied only to sources with a sufficiently large  number of detected maser features.
For  WB89-234 and IRAS~23139+5939, a formal solution was found for both the Keplerian disk and the conical flow model, but only this latter provided a physically reasonable solution (as it will be described in more details in the following). For Sh~2-255\,IR a minimum was found only for the conical flow model. No model fit was attempted for OMC2 because of the (too) low number (three) of features detected.

%[An analysis  based on the physical plausibility of the solutions and/or a comparison with previous observational results convinced us that a conical outflow rather than a keplerian disk model could explain  positions and velocities of maser features seen towards WB89-234 and IRAS~23139+5939.]

%In the following  we describe peculiar aspects of the individual sources.

As is evident from  Fig.~\ref{s255_maps} and~\ref{omc2_maps} (lower panels), Sh~2-255 IR and OMC2 show an elongated spatial distribution  of water maser features, which are aligned along a direction close to the orientation of the molecular outflows seen at much larger scales. Pure ``morphological'' considerations might suggest that the   maser features are associated with the inner portion of the jet/outflow systems. However, in this case one would also expect the transversal velocites to be mainly directed  along the jet/outflow axis.
%OMC2
This geometrical condition is observed towards OMC-2, where the single measured (relative)  proper motion is approximately oriented along the elongation axis of the maser feature distribution. 
An edge-on rotating disk interpretation appears to be less probable, since  the disk radius  and the line-of-sight velocity dispersion as deduced by the observed H$_2$O maser distribution would  imply an enclosed mass of only $\approx$~1~M$_\odot$; this would be in conflict with the (IRAS) bolometric luminosity, which indicates the presence of an intermediate- or high-mass YSO (M$\geq$~6~M$_{\odot}$).
% Assuming a diameter of 90 AU (as derived by the maser distribution) and a line-of-sight velocity dispersion of 5 km s$^{-1}$, one would derive an enclosed mass of $\sim$1.3 M$_{\odot}$, corresponding to a ZMAS star with  expected bolometric luminosity of $\sim$1 L$_{\odot}$, well below the value of 1500 L$_{\odot}$ derived from IRAS fluxes. This huge discrepancy could indicate, if the assumption of keplerian motion is correct, that   the water masers sample only a small fraction of the putative disk, so that the derived value of the YSO mass is a lower limit of the true value.

%S255
On the contrary, towards Sh~2-255 IR,  all the measured {\em relative} proper motions are approximately perpendicular to the outflow axis. 
 In order to obtain for all the maser features absolute velocities directed along the large scale molecular outflow axis, the reference maser feature should move significantly faster ($\geq 100 $ km s$^{-1}$) than the measured relative proper motions.
The best fit solution (found including line-of-sight and transversal velocities)  for  the conical outflow has a a large opening angle, with the cone axis  nearly perpendicular to the line-of-sight.
The orientation of the fitted cone axis onto the plane of the sky is quite different from that of the large scale  ``Br $\gamma$/H$_2$ jet''.
Differences in orientation between the outflow structures on large ($>$ 10000~AU) and small ($\leq$500~AU) scales might be due to either  density gradients in the ambient medium, causing large-angle bends of the protostellar jets, or the presence of multiple, small-scale outflows, whose merging creates the large scale flow \citep{Beu02c,Beu03}.  

%WB234 and IRAS2313
%The spatial distribution of maser features detected towards IRAS~23139+5939 and WB89-234 do not show any preferential direction of elongation.

Towards IRAS~23139+5939,  the CO outflow is nearly oriented along the line-of-sight and not spatially resolved on the plane of the sky. 
Since  water maser features are seen projected onto the plane of the sky, the clustered distribution observed in this source (Fig.~\ref{i2313_maps}, lower panel) might result from a structure elongated along the  line-of-sight, as it would be the case if  maser emission were  associated with the large-scale molecular outflow. This interpretation is consistent with the measured proper motions,  suggestive of a general expansion.
Accordingly, the best-fit conical solution has an axis oriented at close angle from  the line-of-sight, with the vertex (YSO position) close to the geometrical centre of the circular cluster of features.

Towards WB89-234, all the measured proper motions have similar orientation, close to the direction of the molecular outflow, which might suggest  that  the measured, {\it relative}, transverse velocities represent well the ``true'' motion of the maser features, driven by the outflowing gas.
The conical fit result supports these qualitative considerations, the axis of the best-fit cone having a projection onto the plane of the sky close to the position angle  of the molecular outflow.
 %Both sources show  a good agreement of the line-of-sight velocity dispersion of the molecular outflow and the water masers.
%For both IRAS 23139+5939 and  Sh~2-255 IR, the fit is better including a rotational component.
%Il modulo della comp. di rotaz. {\`e} di circa la met{\`a} di quella di espans. 
%[Also for  these two sources, the outflow conical model found the same fit solution when the \ $\chi^{2}$ \ is calculated including also the measured proper motions.] 

For IRAS~23139+5939 and WB89-234, a solution is found also for the Keplerian disk model but the fit of the measured velocities is worse than that for the outflow model.
In addition, in both cases,  the best solution  has been found in correspondence of an edge-on disk, whose geometry is not supported by the observed spatial distribution of maser features that shows no preferential direction of elongation. The fitted disks have  too large diameters ($\geq$~10$^4$~AU) and are very poorly sampled by the detected maser features.
In conclusion, we deduce that, towards both sources, a Keplerian disk interpretation is less likely than the proposed outflow model in accounting for the positions and velocities of the detected water maser features.

This analysis has revealed that the kinematics of water maser features  observed towards  Sh~2-255 IR, IRAS~23139+5939, and WB89-234 may be described by a conical outflow model. 
%This result suggests that in the three sources maser clusters are expanding (supporting the wind interpretation)  rather than rotating  (in agreement with the disk model).
It is  interesting to note that in each case (see Table~4) large opening angles of the conical jets are derived, supporting  the case of large-angle winds. As outlined in Sect.\,1, this result may be of particular interest to derive information on the evolutionary stage of a high-mass YSO and its consequences will be discussed in more detail in Sect.~4.3. 
\begin{table*}
\centering
\begin{tabular}{ccccccc}
\multicolumn{7}{c}{\footnotesize {\bf Table 4:} Best fit models results} \\
& & & & & & \\
\hline\hline
\multicolumn{1}{c}{Source} &     \multicolumn{1}{c}{$\alpha_{\rm c}$}  &  \multicolumn{1}{c}{$\delta_{\rm c}$}  & \multicolumn{1}{c}{$P_{\rm c}$} &  \multicolumn{1}{c}{$i_{\rm c}$} & \multicolumn{1}{c}{$\theta_{\rm c}$} & \multicolumn{1}{c}{$a_{\rm c}$}  \\
\multicolumn{1}{c}{name}  & \multicolumn{1}{c}{($''$)} & \multicolumn{1}{c}{($''$)} &\multicolumn{1}{c}{($^\circ$)} &
\multicolumn{1}{c}{($^\circ$)} &  \multicolumn{1}{c}{($^\circ$)} & \multicolumn{1}{c}{(km s$^{-1}$\ arcsec$^{-1}$)}  \\
\hline
& & & & & & \\
Sh~2-255 IR  & 0.006  & 0.019   & 106   & 72  & 77 &  39    \\
IRAS 23139+5939  & 0.025   & 0.025 &  264   & 23 & 78 &  320    \\
WB89-234 & 0.003   & -0.012 &  58   & 143 & 62 &  82   \\
& & & & & & \\
 \hline
\end{tabular}
\begin{flushleft}
{ \footnotesize Note.-- Col.~1 gives the source name, Cols.~2 and ~3 report the (RA and DEC) sky-projected coordinates of the cone vertex, Cols.~4 and ~5 the position angle and the inclination angle with the line-of-sight of the cone axis, Cols.~6 the opening angle of the cone, and Cols.~7 the velocity gradient.}
\end{flushleft}
\end{table*}

%Once established a kinematical plausib. of our models, we investigate on the consequence of these models on the physical environments\ldots\ldots\ldots\ldots 
%______________________________________________________________________________
%\subsection{Comparison of water masers with radio-mm properties}
\subsection{Physical environment of maser regions}

Water masers  are predicted to  arise in the shocked layers of gas behind both high-velocity ($\geq$~50~km~s$^{-1}$) dissociative (J-type) \citep{Eli89} and slow ($\leq$~50~km~s$^{-1}$) non-dissociative (C-type)  \citep{Kau96} shocks, propagating in dense regions (pre-shock density $\geq 10^7$~cm$^{-3}$).

If water masers originate in expanding jets (as suggested by the results discussed in the previous section), then the {\it absolute} velocity of a maser feature should indicate a lower limit for the shock velocity.
In general,  model fits reproduce  the (well-measured) line-of-sight velocities better than the transversal ones, that are affected by larger errors. Then, to estimate the order of magnitude of absolute velocities on the plane of the sky, instead of using the values predicted by the models, for each component (along the RA and DEC axis) of the (relative) transversal velocity we have calculated the minimum and maximum variation  from the mean value.
The derived absolute velocities are found to be of the order of  8--21~km~s$^{-1}$ (for Sh~2-255 IR), 26--103~km~s$^{-1}$ (for IRAS 23139+5939) and 4--33~km~s$^{-1}$ (for WB89-234). 
These velocities are much larger than the  sound speed (v$_s \leq $~1~km~s$^{-1}$) of the dense ($n_{H_2} \geq  10^5$~cm$^{-3}$) ambient medium through which the outflow is expected to propagate, 
%so the  origin of water masers is naturally explained by means of shock excitation. 
so that shock waves appears as the natural explanation for the excitation of  water masers.
In particular, our results suggest that both C- and J-shocks could play a role in the excitation of  water masers associated with jets, whereas  the presence of extremely high velocity ($>$~100~km~s$^{-1}$) shocks is not indicated by the velocity distribution of the observed water masers (varying in the range from $\sim$1 to $\sim$100~km~s$^{-1}$ for the three sources).

Comparing the line-of-sight with the transversal velocities, one notes that for most of the maser features the relative transversal velocity is in general significantly larger than the corresponding   line-of-sight velocity corrected for the molecular cloud LSR velocity. Interpreting maser features as shocks requires  large differences in amplification gains  along and across the direction of  the shock motion. In particular, velocity gradients are likely to be the largest along the direction of motion, while the best velocity coherence with the longest gain paths should be encountered perpendicularly to the  direction of motion \citep{Eli89}.
Since our EVN observations are sensitive only to the strongest features ($\geq $~0.4~Jy~beam$^{-1}$), the observational result of transversal velocities larger than line-of-sight velocities might reflect the fact that we are observing preferentially shocks moving close to the plane of the sky.

Alternatively, if water masers originate on the surface of an accretion disk (as proposed for AFGL 5142-Group I, Paper I), they would feel the influence of the gravitational field of the massive YSO. 
At distances less than 10--15~AU from the high-mass YSO, models of accretion disks predict sufficiently high values of temperature ($> 400$ K) and density ($> 10^8$ cm$^{-3}$) to match the physical conditions required for the excitation of the water masers. However, at larger distances, shock waves propagating through the disk might be required to warm and compress the masing gas.
 In this case, the pattern of Keplerian motion can be maintained as long as the  shock velocity remains negligible compared to the Keplerian velocity.
For the source AFGL 5142, the Keplerian disk model (which in this case also reproduces  the transversal velocities well) gives  absolute velocities of water masers in the range 7--32~km~s$^{-1}$, corresponding to a range of disk radii from 1 to 400~AU.
A lower limit to the shock velocity can be estimated calculating the sound speed in an accretion disk around a massive YSO. The temperature as a function of the distance from the YSO can be estimated  (in the hypothesis of heating by viscous dissipation) following the calculation by \citet{Nat00}.
The theoretical work by \citet{Pal92} allows to estimate the mass and the radius of a massive proto-star (in this case, 15~M$_{\odot}$ and 4.7~R$_{\odot}$), assuming an accretion rate of 10$^{-4}$~M$_{\odot}$~yr$^{-1}$. The derived temperature of the disk varies from 20 to 1600 K when reducing the distance from the proto-star from 400 to 1 AU, respectively.
This temperature range translates into a variation of the sound speed within the range 0.3--2.4~km~s$^{-1}$. This indicates that, if moderate velocity shocks (Mach number $\leq$~10) propagating throughout the disk are responsible for the maser excitation, the pattern of Keplerian velocities can be maintained.

%Hence, the kinematical analysis presented in this paper and in Paper I to explain the multi-epoch measurements of H$_2$O masers in 4 candidates as high-mass YSOs are compatible with the shock-excited nature of water maser emission.

%RADIO CONT.
Towards each of the observed sources (with the exception of WB89-234), a radio continuum source  is detected.
 Traditionally,  the detection of a radio continuum source has been taken as a proof of the ignition of the nuclear H burning reactions in the nucleus of the forming star:  the heated photosphere would supply the   UV radiation field needed to create an ionized envelope, which consequently would emit free-free radiation at radio frequencies. 
However, since the water maser emission is likely excited by shock waves, the shocks themselves might also account for the radio continuum emission detected towards the YSOs. In this interpretation, the free-free emission would be produced by the shocked, collisionally ionized gas.

Following the theoretical work of \citet{Gha98}, assuming a pre-shock density greater than $10^7$~cm$^{-3}$ (as suggested by the model of \citealt{Eli89})  and a shock velocity in the range 5--100~km~s$^{-1}$ (as estimated in the  previous analysis), one would  estimate an expected radio flux density of the order of 1~mJy, which agrees with the fluxes observed towards our sources.
At the same time, this model  allows to constrain the pre-shock density, shock velocity, and viewing angle to the source knowing the (resolved) angular size, flux density, and spectral index of a thermal radio source (at frequencies of 1.5, 5, and 15~GHz).
Among the studied YSOs, an estimate of the  spectral index $\alpha$ of the radio continuum source (defined as $S_{\nu} \propto \nu^{\alpha}$) is available only for  AFGL 5142 (observed at 5 GHz by \citealt{Mcc91}; flux $\approx$ 0.5~mJy, spectral index $\approx$~0.6) and OMC2 (observed at 15~GHz by \citealt{Ren96}; flux $\approx $ 2.5~mJy, spectral index $\approx$~0.44).
Using these values,  free-free emission of the order of 1~mJy  is accounted for, with densities and shock velocities of 10$^7$--10$^8$~cm$^{-3}$ and $\leq$~100~km~s$^{-1}$, respectively. These values are in agreement with the densities required by maser excitation models \citep{Eli89,Kau96} and with the measured (and modeled) velocities.

%This crude anlysis demonstrates that the kinematics traced by water masers is compatible  with a scenario where shock-waves would excite both H$_2$O masers and radio continuum emission.

%______________________________________________________________________________
\subsection{Which evolutionary stage(s) trace water masers?}
The standard theory of star formation predicts the formation of an accretion disk, as a consequence of the conservation of the angular momentum of the progenitor  molecular cloud.
%, and  simultaneously invokes the ejection of a bipolar  jet, {\it collimated} along the disk axis, from the star--disk system. 
During the accretion phase, the infalling matter suppresses any incipient stellar wind. Breakout will occur through the channel of weakest resistance, and as long as the gas continues to fall onto the disk rather than directly onto the star, this will be at the rotational poles of the accreting protostar, resulting in the ejection of a bipolar  jet, {\it collimated} along the disk axis.
It is plausible that the opening angle of the bipolar outflow widens with time and becomes nearly isotropic, since the wind gradually feels less resistence as the circumstellar envelope disappears.
Following this ``standard'' paradigm, the occurrence of wide-angle conical jets, such as those proposed to explain the water maser kinematics in the studied YSOs, would indicate an evolved phase of the YSO.

 \citet{Tor03} proposed a different scenario, where {\it non-collimated} outflows may result from the emission of wide-angle winds by the YSO at the beginning of its evolution.  Subsequently, the initially isotropic stellar wind may be channelled into a bipolar form by an anisotropic distribution of matter in the surrounding circumstellar environment, originating the collimated outflows commonly observed on large scales towards SFRs. 
Following this alternative scenario,  the observed YSOs could be in a very early evolutionary stage (possibly in a proto-stellar phase).

In order to discriminate between these alternative schemes, we have tried to derive further information on the evolutionary stages of the studied YSOs  based on the radio, millimeter and infrared continuum emissions.

As pointed out in the previous section, the assumption of ionizing flux from the YSO is not absolutely necessary to explain the observed radio continuum emission:
 it might in principle occur during the  proto-stellar phase, excited by shocks associated to the accretion process and/or induced by stellar winds.
On the other hand, accretion rates as high as $10^{-3}$~M$_{\odot}$ yr$^{-1}$ are expected  to occur in  massive YSOs \citep{Oso99,Beu02b} and such massive infalls could cause the quenching of the UCHII radio emission, preventing its detection  even if the YSO is already on the ZAMS (see e.g. \citealt{Pal02}).
Hence, the presence or the lack of radio continuum emission furnishes
no indication to distinguish between the proto-stellar and the ZAMS phase. 
%[The model of \citet{Gha98} predicts that the free-free spectral index increases from the optically thin values ($~$ --0.1) to optically thick ones ($~$ 2) as the shock velocities and preshock densities rise. In this view, one would expect that the radio continuum emission detected towards AFGL 5142 (with an estimated spectral index of 0.6) arises from a higher-density medium [OR: envelop] than that detected towards OMC2 (spectral index $~ -0.1-0.44$). In the assumption that more evolved YSOs are less embedded in dust and gas envelopes (the  matter of  the original cloud part accretes continuosly onto the proto-star and part is ejected by the action of wind/jets), one would deduce that OMC2 is more evolved than AFGL 5142. In principle, however, this can be due to the different ambient in which stars form (e. g., AFGL 5142 could be more massive than OMC2, so the greater $a$ could reflect simply the fact that the original cloud is more massive and than denser). ]

For low-mass stars,  AWB suggested an  empirical evolutionary classification,  based on the ratio between the bolometric luminosity, $L_{\rm bol}$, and the monocromatic luminosity at 1.3~mm, $L_{\rm mm}$. Since in the proto-stellar phase the bolometric luminosity is a rising function of the stellar mass, while the  luminosity at 1.3~mm is proportional to the mass of the accreting envelope, the ratio  $L_{\rm bol}/L_{\rm mm}$ is thought to provide a relative measure of the mass ratio $M_*/M_{\rm env}$: in this scheme, sources with lower values of this ratio are less evolved.
For example, true protostars (the so called Class 0 objects) should  have  $M_*/M_{\rm env}<1$, corresponding to $L_{\rm bol}/L_{\rm mm} \leq 2 \times 10^4$ (AWB).

Despite bolometric and millimeter luminosities for all our sources but
WB89-234 are available in literature, we believe that the same technique,
applied successfully to low-mass-mass YSOs, cannot be applied to their 
high-mass counterparts, and hence, to our studied objects. In principle, 
one could calculate the $L_{\rm bol}/L_{\rm mm}$ ratio and derive a tentative 
evolutionary pattern for our high-mass YSOs to determine if they are in 
a protostellar or ZAMS phase. However, the basic assumption made by 
AWB that $L_{\rm bol}\propto M_*$ (i.e. that the object luminosity derives only from the conversion of its gravitational energy) is valid only if we are 
dealing with "true" protostellar objects, that is precisely the 
information we are seeking for. For a ZAMS star the mass-luminosity 
relation would be totally different rendering the method by AWB not applicable.

Even if  sensitive and high resolution observations of a high-mass YSO were available providing  accurate values of its radio, millimeter, and infrared luminosities (allowing to disentangle its emission from contributions of other  closeby YSOs), these data alone would not suffice to establish  its evolutionary stage.
%%%%%i.e., whether it is spending a  protostellar or a ZAMS phase. 
A possible way to face the problem is complementing the continuum data with high-angular resolution, thermal or maser line  observations, in order to derive information on the kinematics of the gas close to the YSO. 
In the case that the measured pattern of motion  were compatible with rotation and/or contraction in a keplerian disk, one might derive the dinamical mass of the central object, M$_*$. Then, using the accreting envelope mass, M$_{\rm env}$, estimable through  millimeter continuum data, one might calculate the  $M_*/M_{\rm env}$ ratio, which would provide a direct information on the evolutionary stage of the YSO.

%__________________________________________________________________
%
\section{Conclusions}

Using the EVN we have conducted  VLBI  observations of the 22.2~GHz water masers towards four high-mass SFRs (Sh~2-255 IR, IRAS~23139+5939, WB89-234, OMC2) for three epochs (from June to November 1997). 
The water maser emission likely originates close (within hundreds of AU) to a forming high-mass YSO.
Several distinct maser features (on average $\approx$ 10) have been detected for each source and, for those persistent over three epochs, proper motions are derived.
The amplitudes of the proper motions are generally found to be larger than the range of variation of the corresponding line-of-sight velocities. For each source, the proper motion orientation is indicative of expansion.
The LSR velocity variation of water maser features agrees with the line-of-sight velocity dispersion of the molecular outflows detected towards the four high-mass YSOs.

Three different kinematic  models, i.e. a spherical expanding shell, a Keplerian rotating disk, and a conical outflow have been fitted to the 3-dimensional velocity field of the detected maser features. The results of the model fits  suggest that the water maser features are most likely  tracing the inner portion of the  molecular outflows detected at much larger scales. In each case large opening angles of the conical jets are derived.

 The gas kinematics as traced by the observed water masers is compatible with the prediction that in star-forming regions the 22.2~GHz H$_2$O  masers arise behind shock waves. The shocks themselves may also account  for the excitation of the radio continuum sources  detected towards 3 out of 4 of the studied high-mass YSOs.

Basing on current theories of star formation and recent observational evidences of outflow collimation, we believe that the wide-angles 
derived for the 
conical-jets in our YSOs indicate a particular evolutionary stage, 
either evolved or extremely young. However, from the data available at present it is impossible to confidently discriminate between the two scenarios and trace an evolutionary pattern for our high-mass YSOs.

The fact that only 5-6 antennas could take part in each of our EVN observations is reflected in a limited channel map sensitivity threshold ($\approx$ 0.3 Jy/chan). Hence, we are planning to carry out higher-sensitivity multi-epoch VLBA observations of this sample of  YSOs, in order to better constrain the kinematic scenario suggested by this work.

\begin{acknowledgements}
      The European VLBI Network is a joint
facility of European, Chinese, South African and other radio astronomy
institutes funded by their national research councils.

We are very grateful to R. Cesaroni  for providing  the VLA map of 22.2~GHz H$_2$O masers for the source Sh~2-255 IR.  
\end{acknowledgements}

\bibliographystyle{aa}
\bibliography{biblio.bib}

\end{document}